# An Open-Source Hardware and Software Toolkit to Enable Agentic RHEED-Guided Thin-Film Synthesis

*Asraful Haque[1]*, Christopher M. Rouleau[1], Rama K. Vasudevan[1], Sumner B. Harris[1]**

Center for Nanophase Materials Sciences, Oak Ridge National Laboratory, Oak Ridge, Tennessee 37831, United States.

*Correspondence should be addressed to: haquea@ornl.gov or harrissb@ornl.gov

## Abstract

Reflection high-energy electron diffraction (RHEED) provides rich information about evolving surfaces during thin-film growth, but non-automated, operator-dependent alignment and fragmented analysis workflows limit its potential in fully autonomous synthesis. Here, we present an open-source hardware and software toolkit that makes RHEED control and quantitative analysis accessible to operators and artificial intelligence (AI) agents. Demonstrated on a pulsed laser deposition system, the toolkit provides programmable electron-optics control, automated beam alignment and rocking-curve acquisition, and a training-free method for crystallographic azimuthal alignment. The auto-RHEED application extracts structural and growth-related observables through shared graphical and programmatic interfaces, including a Model Context Protocol (MCP) server. An agent-driven demonstration shows how natural-language requests can guide the selection, configuration, and execution of quantitative analyses. An extensible adapter interface streamlines the incorporation of community-developed methods for AI analysis and RHEED simulation as they emerge, supported by Markdown implementation guides designed for AI coding agents. These capabilities support complementary descriptions of surface evolution through physical measurements and learned image representations while providing a practical route for incorporating new computational methods. Together, these tools reduce barriers to automated and agentic RHEED measurements and establish a foundation for future agentic control of thin-film synthesis guided by the evolving surface.

KEYWORDS: autonomous synthesis, pulsed laser deposition, reflection high-energy electron diffraction (RHEED), FPGA-based instrument control, automated beam alignment, vision foundation models, agentic AI.

## Introduction

Reflection high-energy electron diffraction (RHEED) has served for decades as the principal *in-situ* probe of epitaxial thin-film growth. Because the electron beam strikes the surface at a grazing angle, the diffraction pattern is sensitive to the first few atomic planes of a crystal surface, about 1 nm in depth. During thin film growth, the intensity of the specular reflection is typically used to provide a measure of the growth mode, with layer-by-layer growth indicated by periodic intensity oscillations.[1-3] The diffraction pattern itself encodes information about the surface morphology, in-plane lattice spacing, and crystal phase. The rich data provided by RHEED and the relatively simple experimental setup have established it as the standard real-time diagnostic for atomically controlled synthesis of oxide thin films and heterostructures with physical vapor deposition (PVD) techniques such as pulsed laser deposition (PLD)[4, 5] and molecular beam epitaxy (MBE).[6]

In parallel, materials synthesis is undergoing a rapid shift toward autonomous experimentation. Self-driving laboratories that couple automated hardware with machine learning (ML)-driven decision making are accelerating discovery across chemistry and materials science.[7-10] Robotic platforms now plan and execute solid-state syntheses of novel inorganic compounds,[11] and large language model (LLM)-based agents have designed, planned, and performed complex experiments.[12,13] Thin-film growth is following this trajectory: our group previously demonstrated an autonomous PLD platform that combines *in-situ* spectroscopy with Bayesian optimization to identify growth regimes of two-dimensional and functional oxide materials[14,15] and Shimizu *et al.* demonstrated autonomous sputtering.[16] Separately, similar non-autonomous optimization schemes have guided MBE[17, 18] and PLD of oxide films[19].

With the community push towards autonomous PVD, RHEED should be the natural real-time feedback signal, yet it remains one of the least automated components of the experimental PVD workflow. Recent studies have applied ML to RHEED for pattern classification and embedding,[20-22] prediction of stoichiometry,[23,24] on-the-fly detection of growth-mode transitions[25], and real-time feedback control of quantum dot density.[26] ML has also been used to monitor substrate deoxidation with real-time feedback[27] and to detect substrate rotation errors from RHEED patterns.[28] These advances depend on reproducible acquisition geometry: the beam position, incidence angle, and sample azimuth must be established and controlled for the intended measurement. In particular, quantitative comparison of patterns acquired along a crystallographic zone axis requires reproducible azimuthal alignment. In practice, these adjustments often rely on an operator judging the diffraction pattern by eye, introducing variability between samples, measurement sessions, and laboratories. Automating these aspects is therefore an important step toward reproducible RHEED measurements and reliable experimental feedback.

Automated identification of crystallographic directions from rotating-substrate RHEED videos has very recently been demonstrated using supervised neural networks trained on labeled datasets for a single material system.[29] To our knowledge, however, no training-free automated procedure for aligning the RHEED beam and the sample azimuth has been reported. Likewise, quantitative analysis of RHEED image sequences is commonly performed with *ad hoc*, instrument-specific scripts that require manual configuration and lack standardized interfaces for automated orchestration; the only integrated commercial analysis software (kSA 400, k-Space Associates) is license-restricted, while open-source efforts target static pattern analysis and kinematic diffraction simulation,[30,31] or provide dynamical calculations as standalone executables that require additional integration into experimental analysis workflows.[32, 33] The bottleneck that prevents RHEED from serving as a machine-readable observable in autonomous synthesis is therefore not only the control hardware but also the acquisition methods and analysis surrounding it. Indeed, to our knowledge, direct programmatic control of the RHEED electron optics themselves has not previously been demonstrated with open-source hardware and software.

The expanding range of machine-learning and computational methods for RHEED also creates an integration challenge: newly developed analyses must be connected to experimental data, visualization, and automated decision-making before they can become useful during synthesis. An extensible analysis platform should therefore provide a common route for incorporating community-developed methods as they emerge. Separating these methods from data handling and instrument-specific interfaces would allow advances in image representation, temporal analysis, and diffraction simulation to become accessible within established experimental workflows.

Here we present complementary hardware and software tools for programmable RHEED acquisition, automated alignment, and quantitative analysis, demonstrated on a RHEED-equipped PLD system. First, we enable control of the electron gun and deflection optics through a custom, low-cost field-programmable gate array (FPGA)-based control chain and calibrate the coupling between the deflection coils to establish an empirical beam-centering condition for alignment and automated rocking-curve measurements. Second, we introduce a training-free routine that evaluates the mirror symmetry of the diffraction pattern to identify the zone-axis alignment automatically. Third, we introduce Auto-RHEED, a browser-based application with an independent Python analysis core that extracts in-plane spacing, apparent coherence length, and growth rate from RHEED image sequences. Its extensible adapter architecture accommodates user-selected AI models, simulation engines, and community-developed analysis methods through defined interfaces, sharing data handling, visualization, and provenance across these capabilities. A graphical interface and a Model Context Protocol (MCP) server make these capabilities accessible to operators, external programs, and AI agents. We demonstrate agentic analysis through an AI agent that responds to a natural-language request by selecting analysis routines, configuring their parameters, and interpreting the resulting measurements. Incorporating community-developed methods through adapters for temporal analysis and dynamical diffraction simulations further demonstrates Auto-RHEED's extensibility. These capabilities support pretrained vision-model inference, temporal analysis, and physics-based diffraction calculations within a common experimental workflow accessible to AI agents. Together, these contributions address distinct barriers to reproducible, quantitative RHEED measurements and provide building blocks for future agentic control of thin-film synthesis.

## Results and Discussion

### Experimental setup and FPGA-based control system

The primary goal of our electron beam control hardware is to enable arbitrary beam steering for automated sample alignment and the collection of RHEED rocking curves (which are used to evaluate surface termination). Figure 1(a) shows our reflection high-energy electron diffraction (RHEED) system integrated with a pulsed laser deposition (PLD) chamber for real-time monitoring of the film surface during growth. A common differentially pumped RHEED setup for PLD consists of an electron gun, magnetic focusing lens, and an aperture produce a narrow beam that is steered by two magnetic deflection stages, each with a vertical (X) and horizontal (Y) coil pairs; X1/Y1 (coil 1) and X2/Y2 (coil 2). Using two stages in series allows the position and incident angle of the beam at the sample to be set independently, enabling automated alignment and controlled rocking of the beam. The beam strikes the sample inside the ultra-high-vacuum (UHV) chamber at a small grazing angle $\alpha$ and the diffracted electrons are detected by imaging a fluorescent screen on a CCD/CMOS camera. The sample can also be rotated about its normal to select the azimuthal angle, which is the crystal direction to which the electron beam is incident. The sample is optimally aligned by varying the azimuthal angle and deflection coils, most commonly by manual adjustment of deflection coil voltages and stepper motor rotation.

To automate RHEED beam controls, we developed the open-source hardware and software control chain shown in Figure 1(b) (also refer to Figure S1). Commands issued by the user, control scripts, or AI agents are sent to an HTTP server running on an FPGA. We implement the FPGA logic on a Xilinx Zynq-based PYNQ platform. The FPGA drives two multi-channel digital-to-analog converters (DAC; Alamada MAXREFDES24), whose analog outputs are grouped into two functions: four channels set the deflector coils (X1, X2, Y1, Y2), and an additional four channels set the electron-beam parameters (emission current, accelerating voltage, focus, and grid). Placing beam steering under FPGA/DAC control in this way allows arbitrary, synchronized voltage sequences to be applied to the coils, which is the enabling capability for general electron beam control and the automated rocking-curve coil-calibration described below.

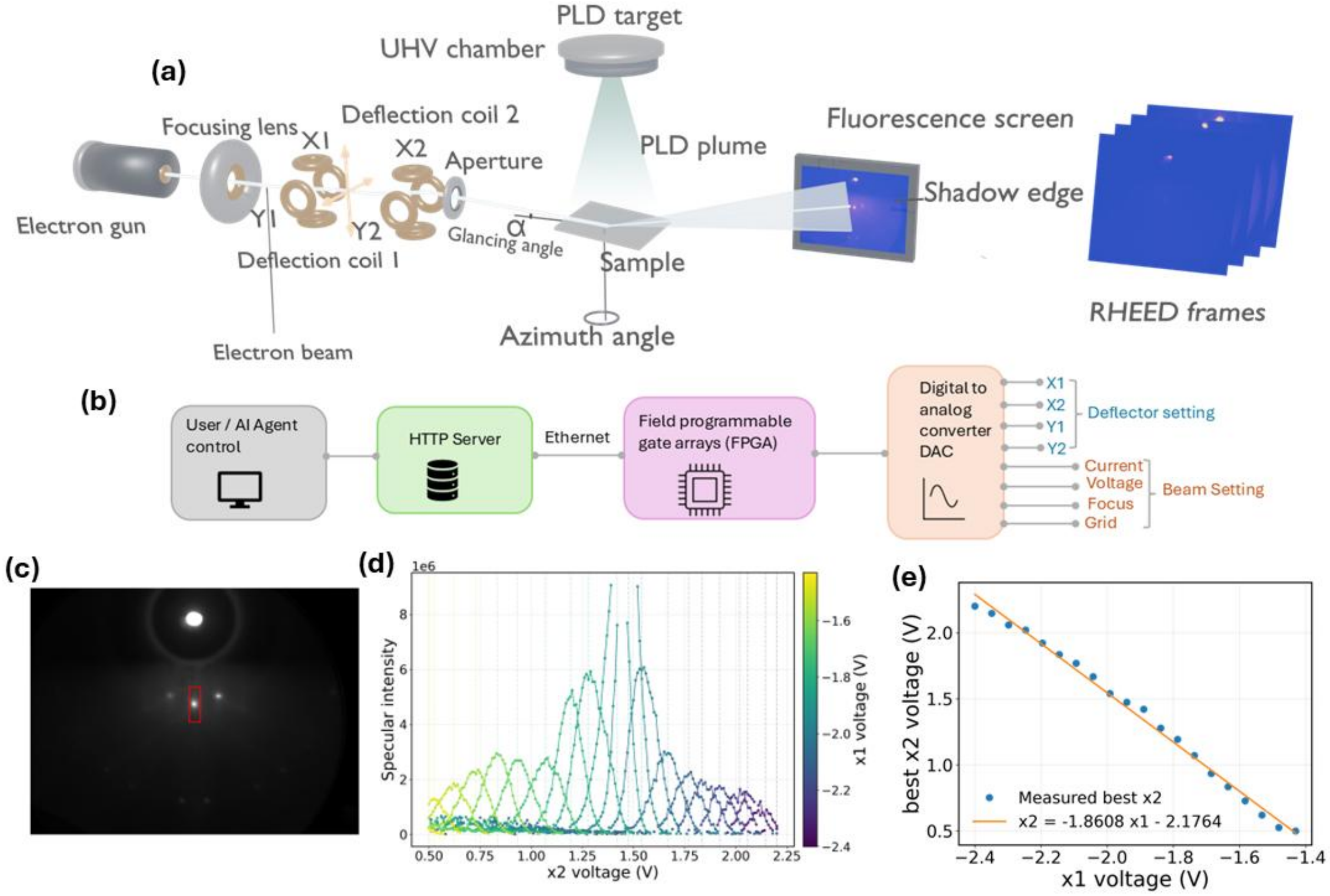

**Figure 1. Automated RHEED beam control and centering calibration on a PLD system.** (a) RHEED-PLD geometry: the electron beam is focused and steered by two deflection stages (X1/Y1, X2/Y2) onto the sample at a grazing angle $\alpha$, and the diffraction pattern on the fluorescence screen is imaged continuously during growth. (b) Digital control chain: commands from the user or Python-based control scripts pass over Ethernet through an HTTP server to an FPGA-driven DAC that sets the four deflector voltages and the electron-beam parameters. (c) Representative RHEED frame; the red box marks the specular region of interest used as the alignment signal. (d) Specular intensity versus X2 at a series of fixed X1 voltages (color scale), each curve showing a single well-defined maximum. (e) Optimal X2 versus X1 extracted from (d); the linear relation X2 = −1.8608 X1 − 2.1764 ($R^2$ = 0.994) defines the beam-centering condition.

Experimentally, collecting a RHEED rocking curve involves sweeping the glancing angle $\alpha$ of the electron beam while maintaining the same position on the sample, which requires coordinated changes in the vertical deflection coil voltages (X1, X2). To increase $\alpha$, X1 steers the beam away from the sample surface and X2 directs the beam back to the sample, and vice versa for decreasing $\alpha$. The relationship between the two coils must therefore be calibrated so that the beam landing position on the substrate remains approximately constant during the angular sweep. The measured RHEED intensity can be expressed as $I(\alpha) \propto I_{diff}(\alpha)G(\alpha)\Phi(\alpha)$, where $I_{diff}(\alpha)$ is the diffraction response, $G(\alpha)$ accounts for geometric effects such as beam position and footprint, and $\Phi(\alpha)$ represents incident electron-beam flux, which can vary if the beam is partially clipped by the aperture at large deflection values. Thus, intensity variations in a rocking curve can arise from both diffraction and non-diffraction effects. During alignment at fixed $\alpha$, where the diffraction condition is nominally unchanged, the deflection-coil relationship is determined by maximizing the specular intensity, thereby reducing intensity losses associated with beam geometry, while variations in electron-beam flux are corrected via direct-beam intensity calibrations.

The procedure to determine the (X1, X2) relationship is described in Figures 1(c)-(e). Figure 1(c) shows a representative RHEED image with the specular reflection tracked by a fixed rectangular region of interest (red box), whose integrated brightness serves as the alignment signal (acquisition and tracking details in Methods). Figure 1(d) shows the specular intensity as a function of X2, with each curve recorded at a fixed X1 (colorbar). Every curve exhibits a single, well-defined maximum whose position shifts monotonically along X2 as X1 is varied, directly visualizing the coupled action of the two coils on the beam footprint. For every X1, the optimal X2 that maximizes the specular intensity ($X2_{max}$) (dashed vertical lines) marks the coil setting for optimal geometric alignment at that particular $\alpha$, where pairs of (X1, $X2_{max}$) each represent an optimally aligned glancing angle $\alpha$.

To extract the "coil constants", we plot $X2_{max}$ against X1. The relationship is linear across the full range, shown in Figure 1(e), and a least-squares fit gives X2 = −1.8608 X1 − 2.1764 ($R^2$ = 0.994). This line defines the locus of coil settings that keep the beam footprint the same across various glancing angles. The high linearity ($R^2$ = 0.994) shows that a single first-order coefficient fully describes the X1-X2 coupling over the working range, providing a simple, robust alignment condition required for rocking curve measurements.

**Automatic azimuthal RHEED alignment using self-symmetry**

Having established automated geometric beam alignment across the accessible range of $\alpha$ (Figure 1), we next address automated azimuthal alignment of the crystal to a symmetric zone axis. In this scenario, we do not employ FPGA beam controls but instead use automated RHEED camera image capture and sample stage motion control which is custom to our PLD chamber. However, this degree of automation is common enough that most PVD systems can be adapted to run the procedure and a simple API to coordinate sample stage rotation with camera acquisition can be written with the help of large language models.

For clarity, we specify that the sample rotation stage provides a laboratory-frame azimuthal angle $\varphi_{lab}$ whose zero is defined by the motor rather than crystallographic orientation. Consequently, the crystallographic azimuth $\varphi$ is related to the stage angle by an offset $\varphi_0$ such that $\varphi = \varphi_{lab} - \varphi_0$. In this example, the procedure is demonstrated on a $SrTiO_3$ (001) substrate, for which the [110] azimuth serves as target alignment direction. Conventionally, alignment is done by eye, the operator rotating the sample until the RHEED pattern "looks symmetric," which is subjective and difficult to reproduce between operators/facilities.

Figure 2 describes a quantitative azimuthal alignment method. As the sample is stepped through $\varphi_{lab}$, we score the left-right mirror symmetry of each RHEED frame and define the optimal azimuthal alignment where $\varphi_{lab}$ produces the maximum diffraction pattern symmetry, and calculate the sample specific offset $\varphi_0$. The complete procedure includes direct-beam and shadow-edge detection, tilt correction, band-pass preprocessing, and the self-symmetry metric calculation (details in Methods). Here, we describe its application and the resulting alignment. Because a genuinely aligned pattern is symmetric only about an axis through the direct beam, the score requires two geometric references: the mirror axis and the pattern tilt, both of which we extract automatically from the first image of the angle scan.

The mirror axis is fixed by the direct beam, which is the brightest feature of the pattern and can be located robustly as the highest-intensity region on the opposite side of the shadow edge (the sample-horizon boundary below the direct beam) from the diffraction features. In this case, its horizontal position defines a vertical symmetry axis. In-plane tilt due to sample mounting rotates the pattern, which breaks mirror symmetry even at perfect azimuthal alignment; we therefore also measure and correct the tilt. The tilt angle is obtained from the shadow edge which is a laterally extended horizontal feature and easily detected with a Sobel image filter. Figure 2(a) shows the Sobel horizontal-edge response of the first frame: the direct beam and a bright horizontal band (the shadow edge) dominate. Integrating the Sobel filter along the horizontal direction (Figure 2(b)) yields a single dominant peak below the beam that identifies the shadow-edge row (orange), clearly separated from the direct-beam response (red dashed). Fitting the edge position across the image gives its slope, from which we recover an in-plane tilt of +1.9° for this dataset. The tilt correction is propagated to every frame so that the shadow edge is horizontal and the mirror axis vertical.

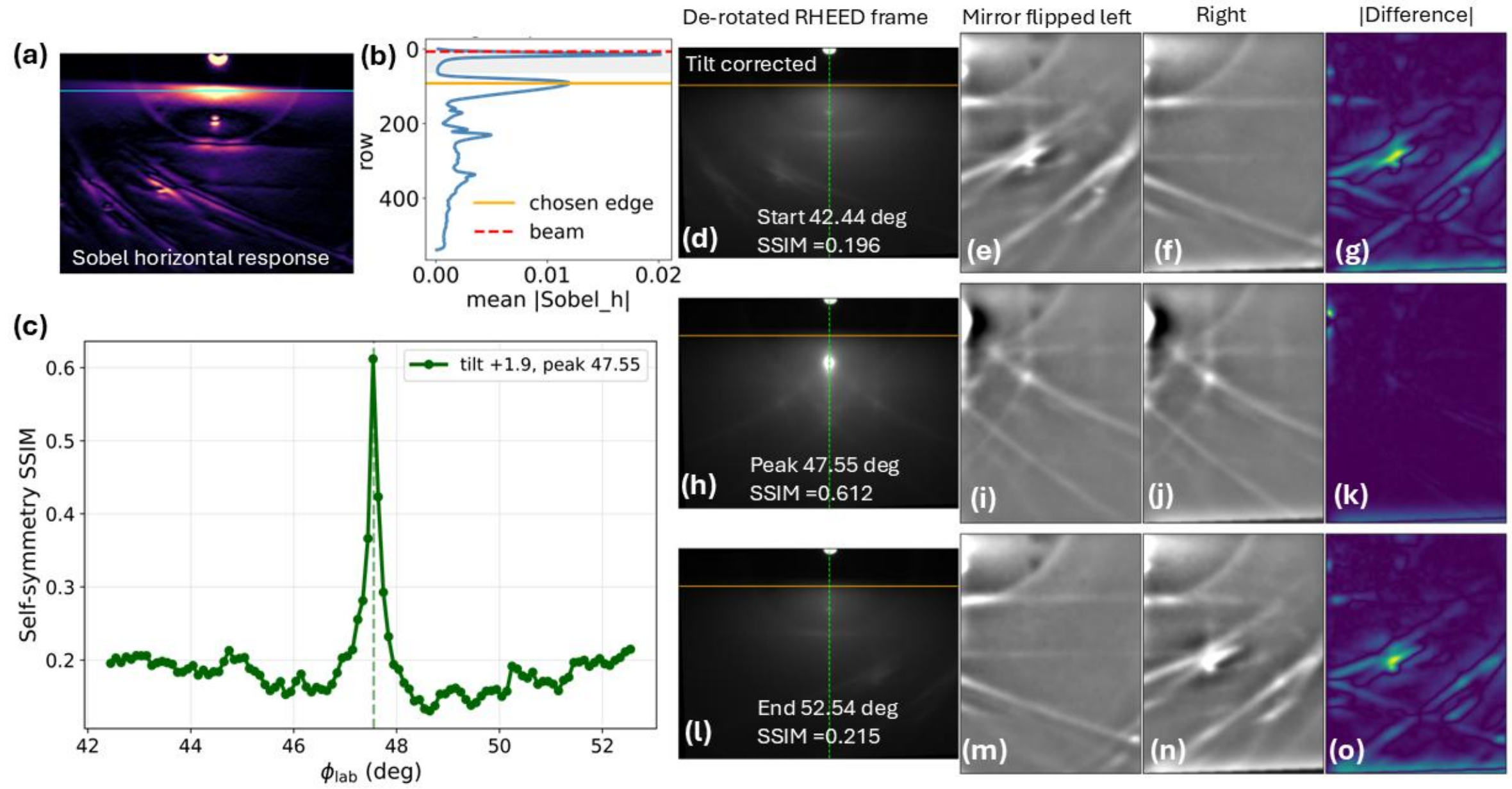

**Figure 2. Automatic azimuthal alignment of the RHEED pattern by self-symmetry.** (a) Horizontal-edge Sobel response is used to locate the shadow edge to correct for sample tilt. (b) Integrated profile of (a); the strongest peak below the direct beam (red dashed) marks the shadow edge (orange), and a linear fit across the image yields the in-plane tilt (+1.9° here), which is removed before scoring. (c) Self-symmetry SSIM versus $\varphi_{lab}$ after tilt correction; the sharp maximum at $\varphi_{lab}$ = 47.55° indicates optimal [110] azimuthal alignment of $\varphi$ = 45°. Tilt-corrected frames with shadow edge (orange) and mirror axis (green) (d, h, l), mirror-flipped left halves (e, i, m), right halves (f, j, n), and absolute differences (g, k, o) at the scan start (42.44°, SSIM = 0.196), the peak (47.55°, SSIM = 0.612), and the scan end (52.54°, SSIM = 0.215).

With the tilt-correction applied, each frame is band-pass filtered with a difference-of-Gaussians kernel (DoG) which removes the smooth background glow from the fluorescent screen and pixel noise while retaining the diffraction streaks and spots. The mirror-symmetry score is then the Structural Similarity Index (SSIM [34]; Eq. (1) in Methods) between the mirror-flipped left half of the pattern and its right half (evaluated over the region below the shadow edge where the azimuth-sensitive diffraction features lie). Figure 2(c) plots this self-symmetry (SSIM) versus $\varphi_{lab}$ which shows a sharp peak at 47.55°, which we define as the optimally aligned [110] azimuth $\varphi$ = 45.00° ($\varphi_0$ = 2.55° offset).

Panels Figure 2(d-o) detail the SSIM response at three representative azimuths: the scan start, the SSIM peak, and the scan end. For each, we display the tilt-corrected RHEED image Figure 2(d, h, l) with the shadow edge (orange) and mirror axis (green), the mirror-flipped left half Figure 2(e, i, m), the right half Figure 2(f, j, n), and their absolute difference Figure 2(g, k, o). At the peak, Figure 2(h-k), the flipped-left and right halves are nearly indistinguishable, corresponding to the maximum SSIM of 0.612. Away from alignment (Figure 2(d-g), Figure 2(l-o)) the Kikuchi bands contribute to asymmetry about the axis, resulting in lower SSIM (0.196 and 0.215). We note that the SSIM does not fall to zero at the scan extremes: there the pattern becomes diffuse and feature-poor, and a smooth image is trivially near-symmetric, which raises the baseline.

The method requires no manual region or feature selection: the beam, shadow edge, tilt, and analysis region are all automatically derived from the data, so the same routine applies unchanged across different substrates and azimuths. Some recent work uses ML algorithms to predict the azimuthal angle but this is not generalizable to any substrate without collecting prior data and is subject to error due to geometry differences between facilities.[29] The only prior input is an approximate $\varphi_{lab}$ to act as the center of the scan range. Because samples can be and are generally mounted in the same orientation, $\varphi_0$ is coarsely known and is mostly instrument specific. The exact $\varphi_0$, which shifts slightly with each new sample mounting, is located automatically by the scan. Importantly, by measuring the tilt from the shadow edge and correcting it before scoring, the alignment is robust to sample tilt, which would otherwise suppress the self-symmetry SSIM peak. Basing the metric on the symmetry of the entire diffraction pattern, rather than on tracking a single specular or diffracted spot, makes it resilient to the spot-detection failures that occur when the specular reflection is weak or ambiguous. Thus, it is ideal to include this alignment routine as part of a broader automated synthesis recipe routine and is generally applicable to different substrate choices and RHEED geometries.

**Auto-RHEED: open-source analysis software for operators and agents**

Beyond electron beam control hardware and generalizable algorithms for sample alignment, the next gap is the availability of open-source, quantitative analysis software that can be operated by both humans and AI. Towards this, we developed a browser-based application called Auto-RHEED[35] for interactive, automated, quantitative analysis of RHEED data. The application is built on a decoupled Python analysis core that can be driven through its graphical interface or through a Model Context Protocol (MCP) server. MCP is an open standard for exposing tools to large language models, so every measurement in the core is advertised to an agent with its name, arguments, and return schema, and is callable without glue code written for a particular model or client; an operator in the browser and an agent over MCP reach identical implementations of the same routine, each in its own analysis session. Figure 3(a) summarizes the analysis pipeline as four stages: (1) raw-data ingestion, (2) geometric calibration, (3) automated per-frame measurement/analysis, and (4) structured export. Auto-RHEED ingests the formats commonly produced by RHEED cameras and acquisition scripts (HDF5 files, standard video in .mp4/.avi/.mov, NumPy arrays, folders of single frames, and raw .img/TIFF detector frames) and synchronizes them to a video viewer and playback timeline. Below we first demonstrate the pipeline from the graphical interface, extracting the in-plane lattice spacing, apparent coherence length, and growth rate from a single growth example (Figure 3), and then show the same routines invoked by an AI agent over MCP, which selects its own regions of interest and returns the analysis from a natural-language request alone (Figure 4).

Auto-RHEED also runs arbitrary AI model inference and diffraction simulations. Because the AI models and physics codes applied to RHEED can evolve rapidly, these are not built into the application but provided as interchangeable adapters, in three families: per-frame inference models, post-hoc analyses of a completed embedding sequence, and diffraction simulation backends. Each adapter declares a typed descriptor giving its name, task, runtime requirements, and default options. Simulation adapters additionally declare their required inputs, capabilities, and an optional JSON schema for their parameters, from which

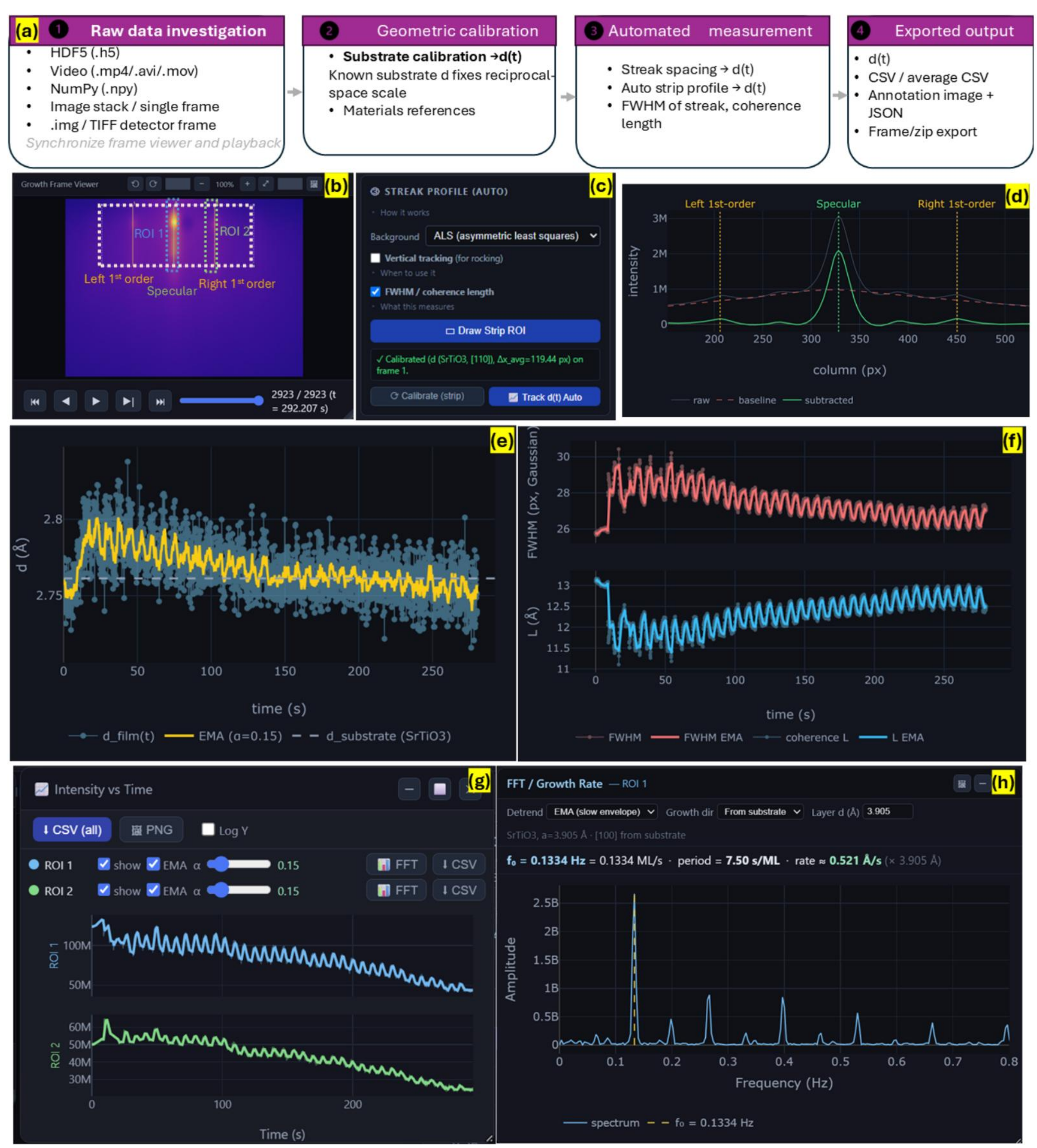

**Figure 3**. **Quantitative structural analysis of $SrTiO_3$ growth in Auto-RHEED**. (a) Four-stage analysis pipeline: data ingestion, substrate calibration, per-frame measurement, and export. (b) Frame viewer with the strip ROI and tracked specular/first-order streaks (ROI 1/ROI 2). (c) Streak-profile controls after calibration along $SrTiO_3$ [110]. (d) Integrated profile: raw signal, baseline, and background-subtracted with automatically detected peaks. (e) In-plane lattice spacing $d$(t) over 2923 frames (grey), with an exponential moving average (yellow) and substrate reference ($\frac{a}{\sqrt{2}} \approx$ 2.76 Å, dashed). (f) Specular-streak FWHM (top) and apparent in-plane coherence length $L$ (bottom). (g) Intensity vs. time for both ROIs. (h) Fourier spectrum of ROI 1. The fundamental $f_0$ = 0.1334 Hz and assuming one monolayer per oscillation gives a deposition rate of the browser controls are generated, so adding a simulator requires no interface code. Adapters are discovered through Python entry points, so an external package registers itself without modification to the core.

Because adapters own their own imports, heavy runtimes are loaded only when a run begins, which keeps PyTorch, Transformers, Ultralytics, and the simulation solver out of the base installation. When GPU inference or simulation is needed, the application can be launched on a remote workstation or high-performance computing cluster and reached in the same browser. Every run records input file hashes, resolved device, pinned package revision, and solver or preprocessing settings, so a reported result can be traced to the exact code, inputs, and parameters that produced it. Four adapters are distributed with Auto-RHEED and used for the analyses described below: DINOv3 embedding, RHAAPsody changepoint detection[25], YOLO instance segmentation, and a bridge to torch-rheed, our PyTorch translation of a Fortran dynamical diffraction code,[32] which makes the same calculations importable as a Python library and executable on a GPU. Each installs its runtime as an optional extra. The interface each family must implement — required methods, result schemas, coordinate and provenance invariants, and verification steps — is specified in markdown documents in the repository, written as instructions that an AI coding agent can follow when adding a new adapter.

The image sequence analyzed here was previously recorded during PLD growth of a $Sr_{2x}Ti_{2(1-x)}O_3$ film on a $TiO_2$-terminated $SrTiO_3$ (001) substrate [24]. Quantitative measurement begins with a geometric calibration. On the first frame (bare-substrate) the user selects a known substrate reflection, the measured streak separation $\Delta x$ fixes the pixel-to-reciprocal-space scale through the small-angle RHEED relation $d = \frac{\lambda L_C}{\Delta x}$, where λ is the relativistic electron wavelength and $L_C$ the sample-to-screen distance (Eqs. 2-3, Methods). Here the $SrTiO_3$ substrate is calibrated along the [110] azimuth ($\Delta x_{avg}$ = 119.44 px on frame 1), for which the in-plane spacing is $\frac{a}{\sqrt{2}} \approx 2.76$ Å. This calibration fixes the reciprocal-space scale used by all downstream measurements (the "substrate-calibration" route in Figure 3(a) stage 2). To follow the in-plane spacing during growth, a strip region of interest (ROI) spanning the specular spot and the two first-order streaks is summed column-wise into a one-dimensional intensity profile for each frame (Figure 3(b) controls; Figure 3(d) profile). The broad, diffuse background is removed (asymmetric least squares is shown, with several estimators available; see Methods), and the specular and $\pm 1^{st}$-order peaks are detected automatically. Averaging the two specular-to-first-order distances gives $\Delta x$ per frame (123.88 px in the frame shown in Figure 3(d)), which the calibration converts to an in-plane lattice spacing. An optional vertical-tracking mode follows the specular row frame-by-frame for rocking or angle-drift geometries.

Applied across all 2923 frames (≈292 s of growth), this procedure yields the in-plane lattice-parameter trajectory $d$(t) in Figure 3(e). The measurement resolves clear, approximately monolayer-periodic oscillations superimposed on a slow relaxation of $d$ toward the substrate reference value (dashed line, $\frac{a}{\sqrt{2}} \approx 2.76$ Å). From the same per-frame profiles, Auto-RHEED simultaneously fits the width of the specular peak and extracts its full width at half maximum (FWHM; Gaussian-fit and half-max read-off options), Figure 3(f, top), and converts this width into an apparent in-plane coherence length L through the same small angle relation, with the streak width substituted for the streak separation (Eq. (4), Methods; Figure 3(f, bottom)), an indicator of the lateral extent of ordered surface domains. Both quantities exhibit oscillations, caused by the periodic sharpening and broadening of the diffraction features as each layer nucleates and coalesces. Note that the coherence length is a convolution of instrumental broadening and true lateral coherence. As such, relative analysis of the coherence length between the starting substrate and the growing film is required without a known instrumental broadening.

The diffracted intensity itself carries the most direct signature of layer-by-layer growth, and Auto-RHEED extracts it in the same session. Figure 3(a) also shows the frame viewer with two rectangular regions of interest placed on distinct diffraction features, the specular spot (ROI 1) and an adjacent diffracted streak (ROI 2), each of which is integrated frame-by-frame to give an intensity transient. The two transients are plotted as vertically stacked, time-synchronized panels in Figure 3(g), where both exhibit pronounced oscillations, superimposed on a slow overall decay of intensity, with an exponential-moving-average smoothing ($\alpha$ = 0.15) overlaid on the raw per-frame signal. Because one full intensity oscillation corresponds to the completion of a single atomic layer, these oscillations are the canonical real-time measure of growth rate.

To convert the oscillations into a quantitative rate, Auto-RHEED computes the discrete Fourier transform of each intensity transient (Figure 3(h); Eq. (5), Methods). The slowly varying envelope, which would otherwise dominate the low-frequency spectrum, is first removed by an exponential-moving-average detrend before transformation. The resulting amplitude spectrum for ROI 1 shows a sharp fundamental at $f_0$ = 0.1334 Hz together with its higher harmonics; since each oscillation period marks one deposited monolayer, this fundamental frequency is directly the growth rate, 0.1334 ML $s^{-1}$. Combining $f_0$ with the out-of-plane interlayer spacing derived from the known c-axis orientation (3.905 Å/ML) yields a deposition rate of 0.521 Å $s^{-1}$ (Eq. (6), Methods). The layer thickness used for this conversion is taken automatically from the active substrate calibration but can be overridden for heteroepitaxial films, and a separate Fourier window can be opened for each region of interest so that growth rates measured from different diffraction features can be compared. Because the growth rate is delivered as a single scalar that updates as frames accrue, it is directly usable by an automated controller, for example to terminate a deposition at a target thickness or to hold a constant growth rate. This extends the analysis pipeline of the previous figure from structural monitoring (in-plane lattice spacing, apparent coherence length) to kinetic monitoring (growth rate), all through the same programmatic interface. Finally, every measurement streams to structured outputs (Figure 3(a), stage 4): *d*(t) and coherence data as per-series and averaged CSV, annotated frames with JSON metadata, and publication-quality figures. These AI-ready observables are compatible with automated and agentic workflows, essentially containing a set of engineered features for AI to reason through time-resolved growth state and issue feedback.

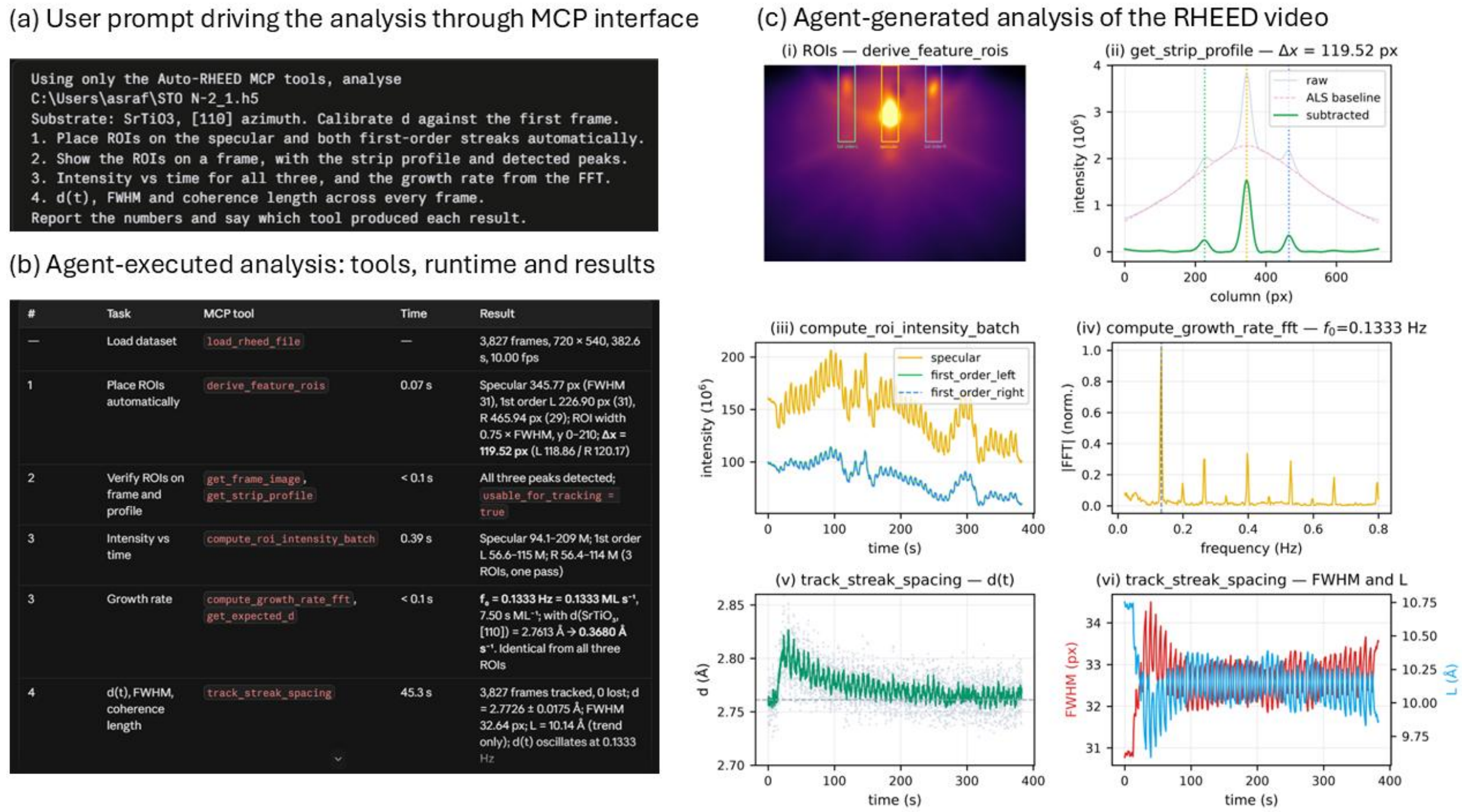

| # | Task | MCP tool | Time | Result |
|---|---|---|---|---|
| — | Load dataset | load_rheed_file | — | 3,827 frames, 720 × 540, 382.6 s, 10.00 fps |
| 1 | Place ROIs automatically | derive_feature_rois | 0.07 s | Specular 345.77 px (FWHM 31), 1st order L 226.90 px (31), R 465.94 px (29); ROI width 0.75 × FWHM, y 0–210; **Δx = 119.52 px** (L 118.86 / R 120.17) |
| 2 | Verify ROIs on frame and profile | get_frame_image, get_strip_profile | < 0.1 s | All three peaks detected; usable_for_tracking = true |
| 3 | Intensity vs time | compute_roi_intensity_batch | 0.39 s | Specular 94.1–209 M; 1st order L 56.6–115 M; R 56.4–114 M (3 ROIs, one pass) |
| 3 | Growth rate | compute_growth_rate_fft, get_expected_d | < 0.1 s | **$f_0$ = 0.1333 Hz = 0.1333 ML $s^{-1}$**, 7.50 s $ML^{-1}$; with d($SrTiO_3$, [110]) = 2.7613 Å → **0.3680 Å $s^{-1}$**. Identical from all three ROIs |
| 4 | d(t), FWHM, coherence length | track_streak_spacing | 45.3 s | 3,827 frames tracked, 0 lost; d = 2.7726 ± 0.0175 Å; FWHM 32.64 px; L = 10.14 Å (trend only); d(t) oscillates at 0.1333 Hz |

**Figure 4. AI Agent-driven RHEED analysis of $SrTiO_3$ film growth through the Auto-RHEED MCP server.** (a) The natural-language prompt; no coordinates, background settings, or calibration constants were supplied. (b) Tools called by the AI Agent, runtimes, and results. (c) Returned analysis: (i) ROIs placed automatically on the specular and both first-order streaks, widths set to 0.75 × each peak's FWHM; (ii) whole-frame strip profile with detected peaks, Δx = 119.52 px; (iii) intensity transients for the three ROIs; (iv) FFT of the specular transient, $f_0$ = 0.1333 Hz = 0.1333 ML $s^{-1}$, giving 0.521 Å $s^{-1}$ using a layer thickness of 3.905 Å; (v, vi) streak tracking over all 3,827 frames, yielding d(t), specular FWHM, and apparent coherence length L trends.

The same routines are exposed through the MCP server, so the analysis of Figure 3 can be requested in natural language rather than assembled by hand. Figure 4 shows a completely agent-driven session on a different $SrTiO_3$ film grown during the same experiment as the one used in Figure 3. The prompt (Figure 4(a)) named the file and asked for automatic ROI placement on the specular and both first-order streaks, intensity transients, a growth rate from the FFT, and per-frame d(t), FWHM, and coherence length; no specific pixel coordinates, ROI widths, background settings, or calibration constants, moving average filter settings, etc. were supplied. The agent correctly inferred the substrate material from the filename, selected and called eight tools in sequence (Figure 4(b)): it located the three peaks, sized each ROI to 0.75 × the fitted FWHM, and then retrieved the annotated frame and strip profile to confirm all three peaks had been detected before committing to the frame-by-frame track, which is the check an operator would normally perform by eye, here carried out by the agent on the returned image. The returned analysis (Figure 4(c)) correctly reproduces similar quantities of Figure 3, giving $\Delta x$ = 119.52 px, a Fourier fundamental of $f_0$ = 0.1333 Hz, and d(t) tracked across all 3,827 frames, the whole session completing in under a minute. The AI agent does not write its own *ad hoc* analysis of the measurement. It calls the same implementations as the graphical interface through the MCP server, so agent- and operator-driven analyses differ only in the parameters chosen, not in the calculation performed. Because those parameters are recorded with every result, an autonomous measurement can be reproduced or audited after the fact. Beyond reproducibility, a natural-language interface requires no prior familiarity with the software, and unattended operation makes it practical to process large numbers of RHEED videos without an expert inspecting each one.

**Automated rocking-curve acquisition and comparison with dynamical diffraction simulations**

Rocking-curve measurements provide an experimental demonstration of the complementary beam-control and simulation capabilities developed here. We use the FPGA-based control chain and empirical deflection-coil calibration to acquire RHEED rocking curves, then compare the measured profiles with dynamical diffraction calculations accessed through Auto-RHEED's torch-rheed adapter. This comparison assesses the consistency of the measured diffraction response with the simulations, providing a physical check on the acquisition and modeling workflow. Because the coils can be rocked under program control (Figure 1), a rocking video is just another sequence for Auto-RHEED, and the glancing angle becomes a measured per-frame quantity. Figure 5 shows a full rocking curve on a stepped STO substrate held at 300 °C with the beam along [100] at 15.8 kV, recorded as 1236 frames at 20 fps over six rocking cycles. Figure 5(a) shows a representative frame showing the direct beam, specular (00) and first-order reflection (01). Intensity-versus-time traces for the specular and the first-order reflections are obtained from Auto-RHEED using rectangular regions of interest and glancing angle was calculated by tracking the direct beam position per frame. Because the beam is clipped at the slit edge near the top of every sweep, the tracked row is fitted with a sinusoid so that the angle remains defined across the clipped stretch. The shadow edge, located automatically by the same routine, fixes $\alpha$ = 0, and the glancing angle of each frame follows from the small-angle geometry $\alpha = \arctan[(r_{edge} - r_{beam}).\frac{s}{L}]$ where s = 7.358x10$^{-3}$ cm/px the screen scale, obtained by fitting a circle to the rim of the phosphor screen of known diameter 4.98 cm (Figure S2), and L=10.5 cm the substrate-to-screen distance. Converting the traces from time to angle in this way yields the rocking curve Figure 5c, which is compared with the dynamical calculations performed with torch-rheed[36]; and a frame from the simulated pattern is shown in (Figure 5b). The shaded bands show that the six cycles reproduce one another to 4% of the mean for 00 through its maximum and to about 20% for (01). The secondary peak in the (01) curve at ~ 2.49°, marked by a vertical dotted line, arises where the (01) diffraction spot crosses a Kikuchi band (Figure S3), which the simulation does not account for. Lastly, the maximum angular range of the rocking curve for this specific geometry is ~ 2.6° due to the differential-pumping pinhole aperture on the electron gun.

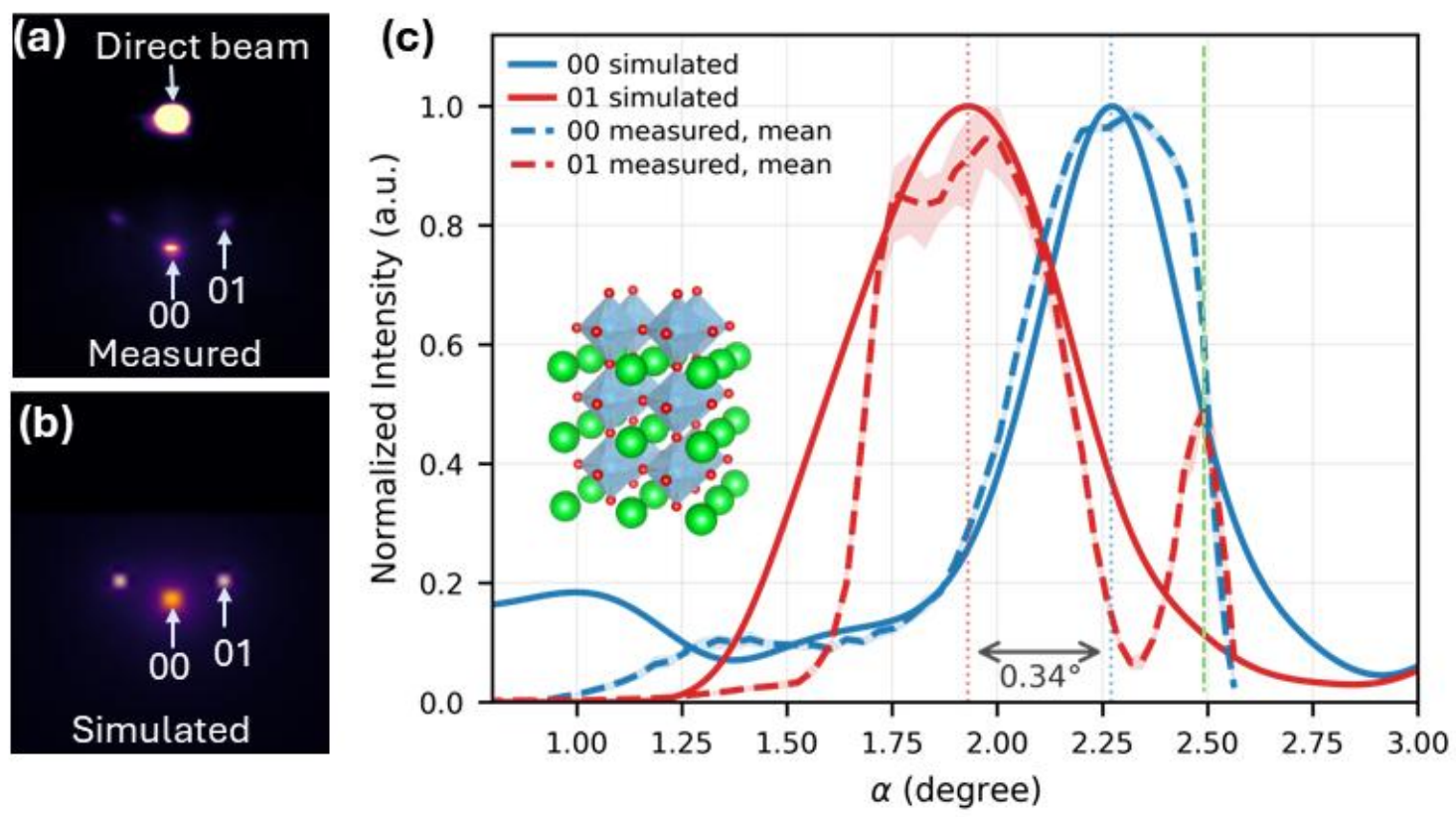

**Figure 5: Rocking curve of $TiO_2$-terminated $SrTiO_3$ (001) along [100] at 15.8kV.** (a) Measured and (b) Simulated RHEED pattern showing the specular (00) and first order (01) reflections. (c) 00 and 01 intensity versus glancing angle α. Dashed lines are the mean of six measured rocking curves, shaded bands show the full range. Solid lines are dynamical simulations convolved with 0.25° FWHM Gaussian to approximate beam divergence and screen broadening. Dotted vertical line mark the peak positions. All curves are normalized with their own maximum. The measured peaks at 2.32° (00) and 1.97° (01) compare with the simulated positions of 2.27° and 1.93° and give a peak separation of 0.35° and 0.34°, respectively. The measured curves terminate at ~ 2.6° because the beam is clipped at the electron gun aperture beyond that angle. Inset: Simulated $TiO_2$-terminated $SrTiO_3$ structure.

### AI inference and community-method integration through adapters

The analysis above converts RHEED into a set of engineered features like in-plane lattice spacing, apparent coherence length, growth rates etc., all of them presuppose a calibration geometry and ROI selection. A complementary route is to let a pretrained vision model describe the diffraction pattern rather than hand selected features. Auto-RHEED exposes two such routes, one requiring no task specific training and one fine-tuned on labelled

RHEED images. In the first approach, each frame of an $La_{0.7}Sr_{0.3}MnO_3$ (LSMO) film grown on a STO substrate was embedded with DINOv3 (ViT-S/16)[37] using the class token, without fine tuning, yielding one 384-dimensional descriptor per frame that is keyed to its native frame index and timestamp and therefore remains synchronized with every classical observable from the same run. Principal component analysis (PCA) of the embedding is shown in (Figure 6(a)) and reveals that the growth traces a single continuous trajectory in embedding space and k-means (with k=4) partitions this trajectory into four successive regimes, which the film occupies in order 0, 3, 2,1 (k-means labels are arbitrary). The representative frames in Figure 6(c) identify them as: cluster 0 covers the bare substrate and the initial nucleation stage, cluster 3 the predominantly streak patterns of 2D growth, cluster 2 the spots superimposed on those streaks that mark the onset of mixed 2D+3D growth, and cluster 1 the spotty pattern with ring segments characteristic of 3D and polycrystalline growth.

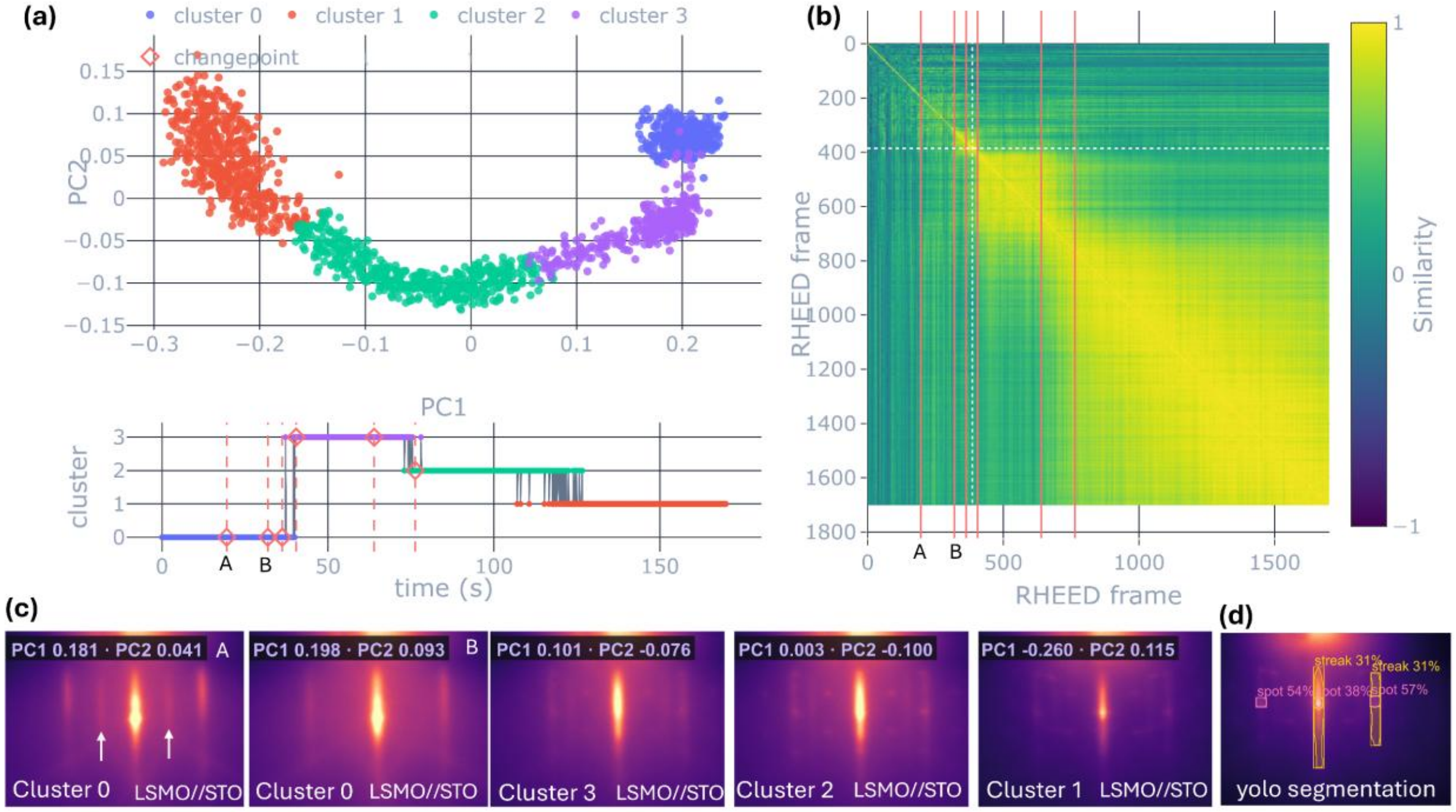

**Figure 6. AI Inference in Auto-RHEED**. (a) PCA of DINOv3 (ViT-S/16) embedding; each point is one frame, colored by k-means cluster label. Top: PCA1-PCA2 projection. Bottom: cluster assignment and PCA cluster versus time, with detected changepoint (red dashed). (b) Changepoints are obtained using an adapter implementing the temporal-analysis approach of RHAAPsody applied here to DINOv3 embeddings. (c) representative frames per cluster with PCA1/PCA2 coordinates. Cluster 0 (frames A and B) covers the substrate and initial nucleation stage, with the arrowed streaks disappearing between the two changepoint; cluster 3 is the predominantly streaked pattern of 2D growth; cluster 2 shows spots superimposed on streaks, marking mixed 2D+3D growth; cluster 1 shows spots with ring segments, characteristics of 3D and polycrystalline growth. (d) Yolo26m instance segmentation of a single frame from the LSMO-STO RHEED video with class labelled and confidence scores.

Figure 6 also demonstrates the incorporation of a community-developed analysis method through the adapter architecture. Kaspar et al. introduced RHAAPsody, which combines pretrained image representations with temporal analyses to identify changes during thin-film deposition.[25] Here, an adapter implements its centered-similarity and changepoint-analysis approach using the DINOv3 embedding sequence as input. The image representation and temporal analysis are supplied by separate adapters, allowing the community method to be applied within Auto-RHEED while retaining a common frame index and time axis.

A similarity center is established from an initial reference period, and the centered cosine-similarity matrix is analyzed to identify changepoints (Figure 6(b)). The first two detections, marked A and B, occur within cluster 0 and correspond to successive changes in the half-order diffraction streaks highlighted in Figure 6(c). These detections complement the broader partition obtained by clustering the embeddings. Their integration illustrates how a published analysis method can be incorporated into the platform and combined with a different pretrained representation within a shared, traceable workflow. Furthermore, a supervised route is available through the same interface; a YOLO instance-segmentation model[38], initialized from pretrained weights and fine-tuned on 108 manually annotated RHEED images (using Meta's SAM3[39] implemented with Ultralytics YOLO26[,40] through custom software[41]) spanning three classes (direct beam, spot, streaks), returns per-instance masks, class assignment and confidence score as shown for a single frame in Figure 6(d).

## Conclusions

This work presents complementary open-source hardware and software tools for RHEED electron-optics control, automated alignment, and quantitative analysis, with programmatic interfaces accessible to operators and AI agents. Programmable beam deflection and an experimentally calibrated coupling between deflection stages enable repeated rocking-curve acquisition, while a training-free self-symmetry metric locates the crystallographic alignment without manually selected diffraction features. Measurements on $TiO_2$-terminated $SrTiO_3$(001) reproduce the principal ordering and separation of dynamical diffraction maxima, providing an experimental check on the acquisition, analysis, and simulation workflow. Auto-RHEED extends these capabilities to time-resolved extraction of in-plane spacing, apparent coherence length, and intensity oscillations, with an agent-driven demonstration showing how these measurements can be automatically assembled from a natural-language request.

These tools support complementary descriptions of the evolving surface. Physically interpretable observables quantify selected structural and kinetic features, while pretrained vision-model embeddings capture broader changes in diffraction appearance. DINOv3 embeddings and temporal analysis distinguish successive pattern regimes during oxide growth without task-specific training, and supervised segmentation provides access to individual diffraction features. Shared analysis routines and recorded parameters make operator- and agent-driven results traceable. The incorporation of RHAAPsody changepoint detection and torch-rheed simulations demonstrates how community-developed methods can become accessible to operators and AI agents through common adapter interfaces. This extensibility allows auto-RHEED to evolve alongside advances in diffraction theory and machine learning methods.

More broadly, connecting instrument configuration, quantitative observation, and computational interpretation provides a foundation for experiments that respond to the evolving film. RHEED can thereby become an actively interrogated source of surface information within automated synthesis, supporting decisions about when to measure, when to adjust growth conditions, and when to stop deposition. By making these capabilities openly accessible, this work creates a practical route toward reproducible, adaptive control of thin-film growth.

## Conflict of Interest

The authors declare no competing interests.

## Author Contributions

A.H.: Data Curation (equal); Investigation (lead); Software (supporting); Writing – original draft (lead); Writing – review & editing (equal). C.M.R.: Investigation (supporting); Methodology (supporting); Writing – review & editing (equal). R.K.V.: Investigation (supporting); Software (supporting); Writing – review & editing (equal). S.B.H.: Conceptualization (lead); Methodology (lead); Investigation (supporting); Data curation (equal); Software (lead); Writing – review & editing (equal).

## Data Availability

The RHEED data that supports the findings of this study are available in supplemental videos. The input files for the torch-rheed simulations are available in the Auto-RHEED repository.

## Code Availability

The code required to reproduce the FPGA-based electron beam controls and HTTP server are openly available at https://github.com/sumner-harris/rheed-fpga.git. Auto-RHEED and its underlying analysis library, the automated-alignment (SSIM) routines are openly available at https://github.com/Asraf235/auto-RHEED.git. The dynamical RHEED simulations are openly available at https://github.com/sumner-harris/torch-rheed.git.

## Methods

### Beam-centering calibration

The specular reflection is tracked with a fixed rectangular region of interest (ROI; 216 × 293 px) positioned on the specular spot, and the integrated ROI intensity serves as the alignment signal. The calibration is a two-dimensional sweep of the vertical deflection coils: X1 is stepped through 20 fixed voltages spanning −2.40 to −1.43 V and, at each X1, X2 is swept over 171 steps with 0.01 V spacing while the ROI intensity is recorded frame-by-frame. For each X1, the optimal X2 is taken as the peak of the intensity-versus-X2 curve, refined to sub-step resolution by a local parabolic fit to the points bracketing the maximum. The resulting (X1, X2) pairs are fit by least squares to obtain the linear beam-centering relation reported in the Results.

### RHEED Alignment

The routine aligns the sample by maximizing the left-right mirror symmetry of the RHEED pattern as the azimuth $\varphi$ is stepped. Beam geometry and tilt are determined once, from the first frame; preprocessing and scoring are applied to every frame. It is implemented in Python (NumPy, SciPy, scikit-image) and requires no manual input.

**Geometry** The direct beam is localized as the brightest connected region in the upper half of the frame (>99.5th intensity percentile); its horizontal position $x_b$ defines the vertical mirror axis, and the beam is masked (radius +5 px) during scoring. The shadow edge below the beam is detected with a horizontal-edge Sobel operator $S_h$ (vertical-derivative kernel) applied to the Gaussian-smoothed (σ = 3 px), min–max-normalized frame. Averaging the response magnitude $|S_h * \hat{I}|$ over central columns ($|x - x_b| \leq 150$ px) gives a row profile whose strongest peak below the beam ($y > y_b + 3r_b$) is the edge row. Fitting the edge position across columns yields its slope m and the in-plane tilt θ = arctan m; each frame is corrected by θ about the beam so that the edge is horizontal and the axis vertical.

**Preprocessing** Each rotation-corrected frame is band-pass filtered by a difference of Gaussians, $J = (G_{\sigma 1} - G_{\sigma 2}) * I$ with σ1 = 3, σ2 = 20 px, which suppresses the smooth background and pixel noise while retaining the diffraction features, and is then standardized to zero mean and unit variance.

**Symmetry score** In the region below the shadow edge, the frame is split at the mirror axis, the left half is mirror-flipped (flip(L)) and compared to the right half (R) by the mean Structural Similarity Index [34] over w × w windows (w = 7):

$$SSIM(a,b) = \frac{(2\mu_a\mu_b + C_1)(2\sigma_{ab} + C_2)}{(\mu_a^2 + \mu_b^2 + C_1)(\sigma_a^2 + \sigma_b^2 + C_2)} \quad (1)$$

where μ, $\sigma^2$, and $\sigma_{ab}$ are the local means, variances, and covariance, C1 = $(0.01L)^2$, C2 = $(0.03L)^2$, and L = 1 (the normalized data range). The frame score is the window-averaged value S($\varphi$) = mean SSIM(flip(L), R), with S = 1 for a perfectly mirror-symmetric pattern.

**Alignment** The aligned azimuth is the maximum of S($\varphi$), refined to sub-step resolution by a local Gaussian fit to the points within ±1.5° of the discrete maximum. Since the geometry is computed once and each frame needs only a rotation, a band-pass filter, and one SSIM evaluation, a full ~100-frame scan is processed in a few seconds.

**Film growth**

The $Sr_{2x}Ti_{2(1-x)}O_3$ films analyzed with Auto-RHEED were previously grown by PLD, fully describes in Harris et al.[24] Described here briefly, sub-monolayer layers of SrO and $TiO_2$ were deposited sequentially by ablating the respective targets with a KrF excimer laser (248 nm, 25 ns; 1.0 J $cm^{-2}$ fluence, 3 Hz repetition rate) at a target-substrate distance of 4.5 cm, a substrate temperature of 700 °C, and 20 mTorr of flowing $O_2$ (base pressure < 3 × $10^{-6}$ Torr). The 5 × 5 mm $SrTiO_3$ (001) substrates (CrysTec GmbH) were annealed at 1050 °C, leached in deionized water, and annealed again at 1050 °C to obtain a stepped, $TiO_2$-terminated surface (Figure S4). RHEED was acquired with a differentially pumped electron gun at 20 kV (Staib TorrRHEED).

The LSMO film (analyzed in Figure 6) was grown on an as-received 5 x 5 mm $SrTiO_3$(001) substrate by PLD. Before deposition, the substrate was ultrasonically cleaned sequentially in acetone, isopropanol, and deionized water for 3 min each. A commercial $La_{0.7}Sr_{0.3}MnO_3$ target was ablated using a KrF excimer laser (248 nm, 25 ns) at 33.7 mJ per pulse, corresponding to a fluence of approximately 1.35 J $cm^{-2}$. The film was grown at 800 °C in 20 mTorr $O_2$ with a 2 Hz repetition rate, 5 cm target-substrate distance. RHEED was acquired with a differentially pumped electron gun at 15.8 kV (Staib TorrRHEED).

**Auto-RHEED analysis**

**Implementation.** Auto-RHEED is implemented as a Python web application: a Flask server exposes a decoupled analysis core (NumPy, SciPy, OpenCV) through both an interactive browser interface (HTML5 canvas overlays, Plotly.js) and an API and MCP server, so identical routines are available to a human operator or an AI agent. Image sequences are read from HDF5, video (.mp4/.avi/.mov via OpenCV), NumPy arrays, image-stack folders, or raw

.img/TIFF detector frames. The application and analysis core are openly available (see Code Availability).

**Electron wavelength and geometric calibration.** The relativistic de Broglie wavelength of the incident electrons is

$$\lambda = \frac{h}{\sqrt{2m_0 eV(1 + \frac{eV}{2m_0 c^2})}} \tag{2}$$

where V is the accelerating voltage. A single calibration converts detector pixels to reciprocal space through the small-angle RHEED relation

$$d = \frac{\lambda L_C}{\Delta x} \tag{3}$$

where Δx is the separation (in calibrated length) between the specular and first-order streaks, $L_C$ is the camera length (sample-to-screen distance), and d is the corresponding real-space in-plane spacing. For the SrTiO3 substrate calibrated along the [110] azimuth, the reference spacing is d = $\frac{a}{\sqrt{2}}$ = 2.761 Å (a = 3.905 Å); this fixes the pixel-to-reciprocal-space scale s (length per pixel) used by all subsequent measurements.

**Strip-profile extraction and in-plane lattice spacing d(t).** For each frame, a user-defined strip region of interest spanning the specular and both first-order streaks is summed along the columns to yield a one-dimensional lateral intensity profile I(x). A slowly varying background is removed by asymmetric least-squares smoothing (Eilers baseline; smoothness $\lambda_{ALS}$ = $10^5$, asymmetry p = 0.01), with rolling-ball, iterative-polynomial, and moving-minimum estimators also available. The specular and ±1$^{st}$order peaks are located on the background-subtracted profile by prominence-based detection and refined to sub-pixel precision by a parabolic fit to each maximum and its two neighbors; frame-to-frame tracking seeds each search from the previous frame's positions. The mean of the two specular-to-first-order distances gives Δx per frame, converted to d(t) via Eq. (3). An optional vertical-tracking mode re-centers the summed band on the specular row each frame to decouple vertical spot motion (e.g. during beam rocking or angle drift) from the lateral measurement. Time series are optionally smoothed with an exponential moving average (EMA), $\hat{y}_t = \alpha y_t + (1-\alpha)\hat{y}_{t-1}$ , where $y_t$ is the raw per frame value at frame t, $\hat{y}_t$ is the smoothed value, $\hat{y}_{t-1}$ is the smoothed value at the previous frame and $\alpha(0,1)$ is the smoothing factor.

**Streak width and apparent coherence length.** Surface order is quantified from the specular-streak width, measured either as a direct half-maximum read-off or from a Gaussian fit (FWHM = 2.3548 σ) of the background-subtracted profile. The width $w$ is converted to an apparent in-plane coherence length L through the same small-angle approximation used for the in-plane lattice spacing, with the streak width $w$ substituted for the streak separation $\Delta x$ in Eq. (3):

$$L = \frac{\lambda L_C}{w \cdot s} \tag{4}$$

ecause L is bounded by the instrument transfer width, its temporal evolution, rather than its absolute magnitude, is the physically meaningful quantity.

**Region-of-interest intensity and growth rate.** Diffracted-intensity transients are obtained by integrating pixel intensity within user-defined circular, rectangular, or line regions of interest across all frames. To extract the growth rate, the slowly varying envelope of a transient is first removed by subtracting an EMA of the signal, $\hat{I}(t)$ (mean- and linear-detrend options are also provided), a Hann window is applied, and the discrete Fourier transform is computed,

$$P(f) = |FFT\{[I(t) - \hat{I}(t)] \cdot w_{Hann}(t)\} \tag{5}$$

with the DC bin set to zero. Here, for simplicity, we assume that one intensity oscillation corresponds to the deposition of one monolayer, the dominant non-zero frequency $f_0$ = argmax $P(f)$ is the growth rate in monolayers per second; the oscillation period is 1/f0 s, and the absolute deposition rate is

$$R = f_0 \cdot d_{layer} \tag{6}$$

where $d_{layer}$ is the out-of-plane interlayer spacing taken from the known crystal orientation c = 3.905 Å or entered manually for heteroepitaxial films.

## Acknowledgements

This work was supported by the Center for Nanophase Materials Sciences (CNMS), which is a US Department of Energy, Office of Science User Facility at Oak Ridge National Laboratory.

## SUPPORTING INFORMATION

Additional experimental details, hardware schematics, calibration procedures, AFM analysis, RHEED image analysis, and supporting RHEED movies are available in Supporting Information.

# Supporting Information for:

# An Open-Source Hardware and Software Toolkit to Enable Agentic RHEED-Guided Thin-Film Synthesis

*Asraful Haque[1*], Christopher M. Rouleau[1], Rama K. Vasudevan[1], Sumner B. Harris[1*]*

1. Center for Nanophase Materials Sciences, Oak Ridge National Laboratory, Oak Ridge, Tennessee 37831, United States.

*Correspondence should be addressed to: haquea@ornl.gov or harrissb@ornl.gov

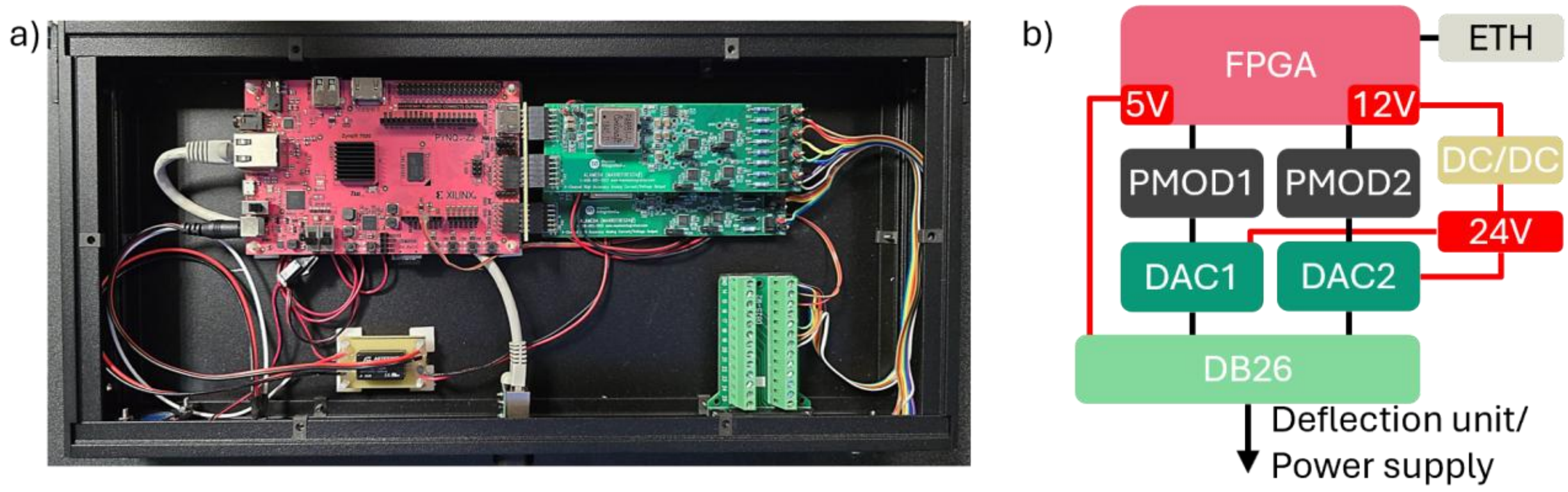

**Figure S1. Custom FPGA-based control unit for the RHEED electron beam control.** (a) Photograph of the assembled 1U rack mounted enclosure, showing the Xilinx Zynq-based PYNQ-Z2 board (left), the two Alameda MAXREFDES24 digital-to-analog converter (DAC) boards mounted to the two peripheral module (PMOD) connectors on the PYNQ-Z2 and the DB26 output connector to the electron deflection and power units (right). (b) Block diagram of the control chain shows ethernet port (ETH) for TCP/IP communication with the PYNZ-Z2 hosted HTTP server, PMOD1 and PMOD2 connections for SPI communication to DAC1 and DAC2. The 24V power input is used to power both DACs, and is stepped down to 12V with a DC/DC to power the FPGA. A 5V input is supplied from the FPGA through the DB26 to switch the electron control units to “computer control” mode (specified in Staib RHEED manual).

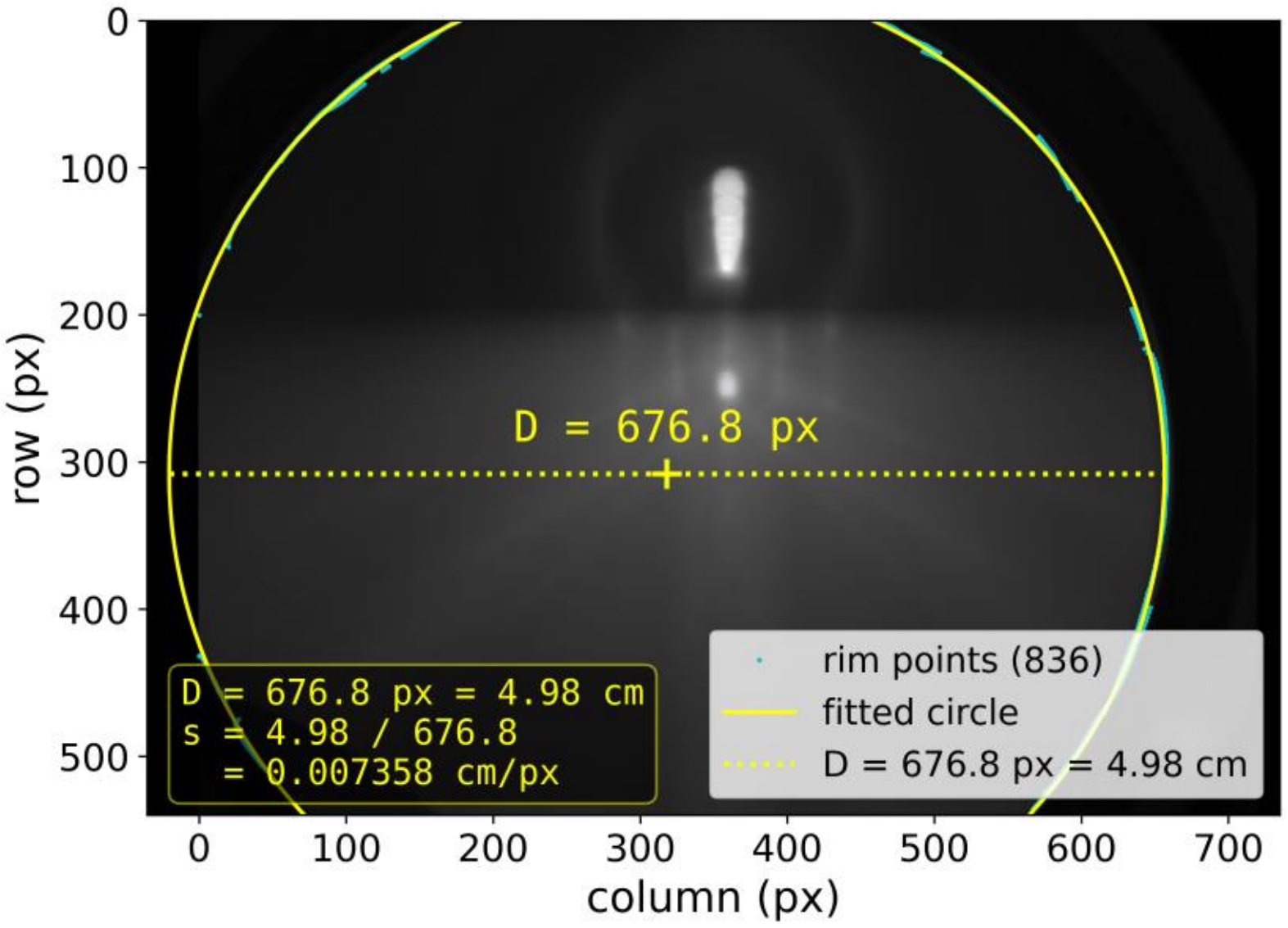

**Figure S2.** Pixel-to-centimeter calibration ($s$) of the phosphor screen ([-100] run, mean of every 5$^{th}$ frame). Cyan points mark the screen edge, located at sub-pixel half-height along every image row and column; the yellow circle is a RANSAC fit to them. The fitted diameter D = 676.8 px (dotted line), the measured screen is 4.98 cm, gives $s$ = 0.007358 cm/px ; the scale used in calculating the glancing angle $\alpha$.

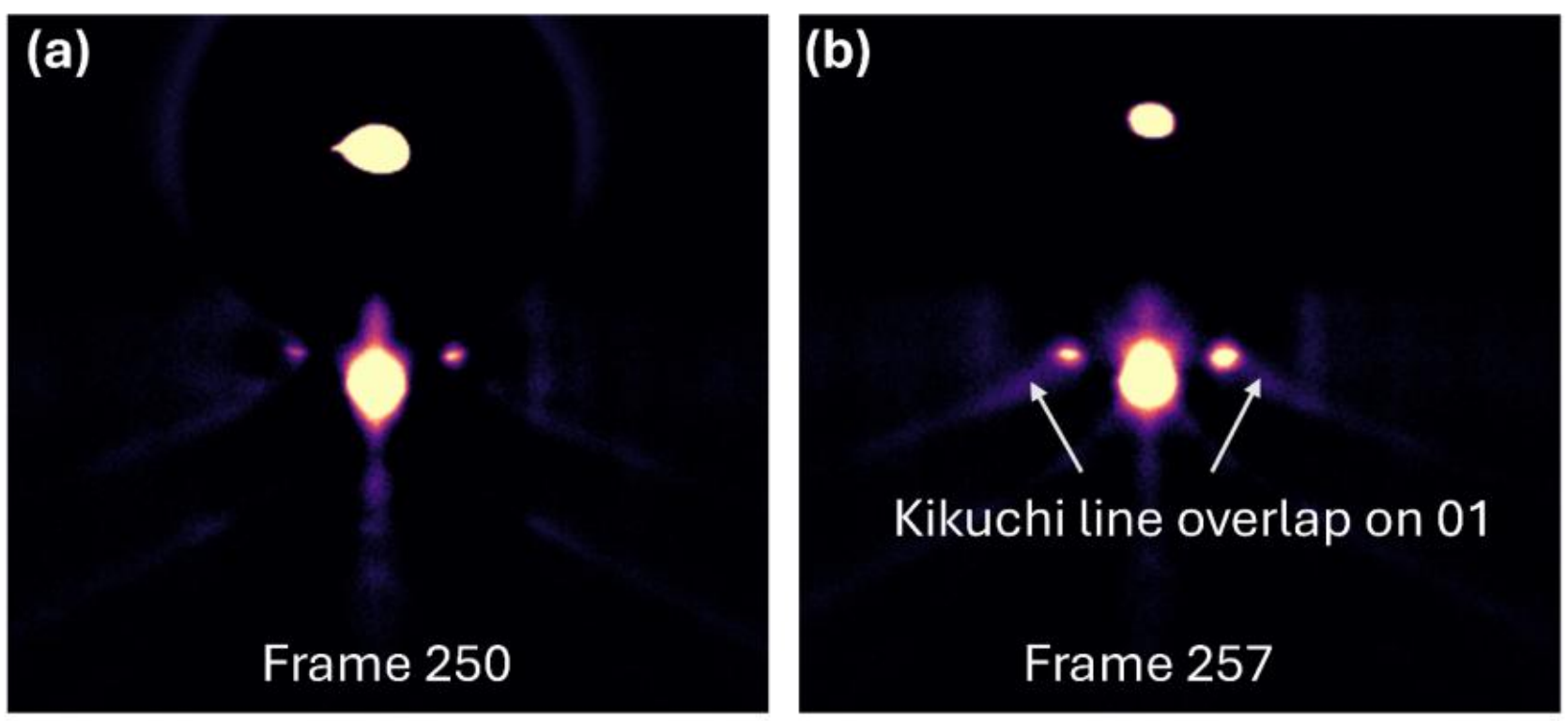

**Figure S3. Origin of secondary 01 maximum at α = 2.49°.** (a) Frame 250 (t=12.50 s, alpha 2.32), at the minimum of the 01-rocking curve, and (b) frame 257 (t=12.85s, alpha=2.49), at the secondary maximum. In (b) a pair of Kikuchi lines has swept onto the first-order spots (arrows), raising their contrast against the background giving rise to secondary peak in Figure 4 (shown by green dotted line).

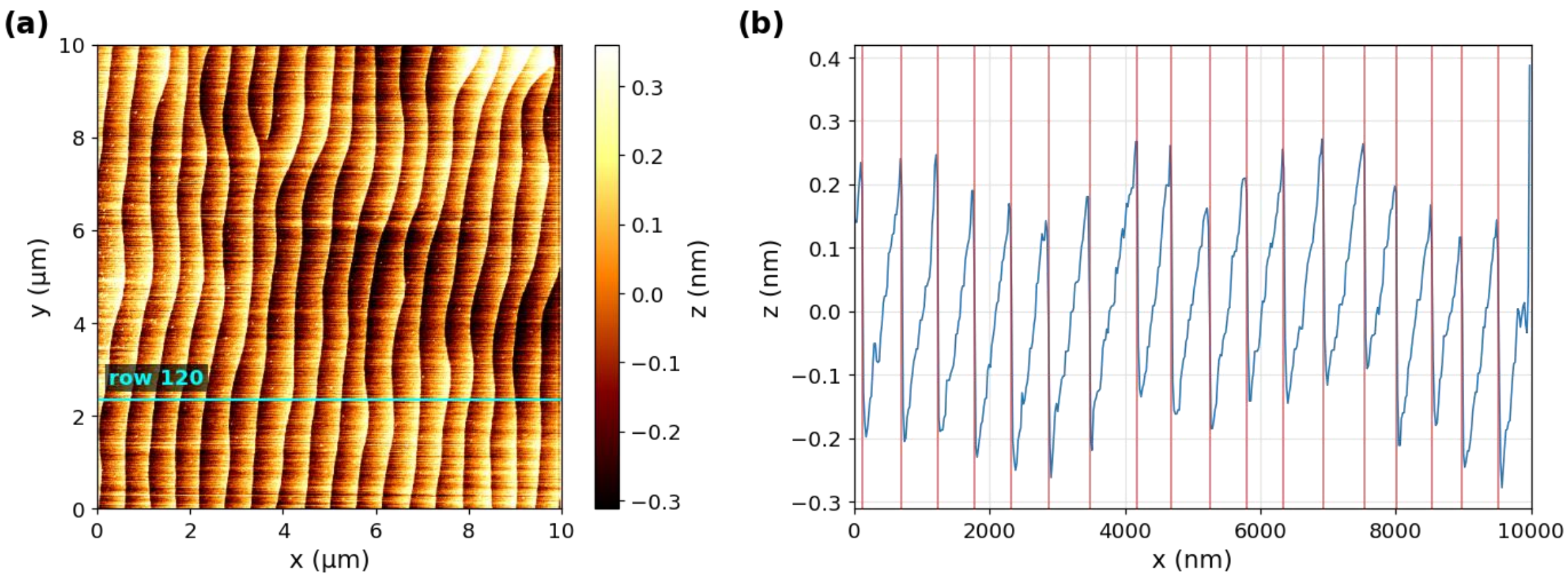

**Figure S4.** Miscut measurement of the $TiO_2$-terminated $SrTiO_3$(001). (a) 10 × 10 μm AFM topography; cyan line = row 120. (b) Profile along row 120 with the 18 detected step edges (red); the sawtooth arises from the per-scan-line plane fit applied during acquisition.

### Supporting Movies/data files

**Movie M1.** STO_stepped_-100_300C_15.8kV_1.4A_5000us.mp4: Rocking curve of $TiO_2$-terminated $SrTiO_3$ (001) at 300 °C, beam along [100], 15.8 kV, 5000 μs exposure, 1236 frames at 20 fps. Six sinusoidal rocking curve cycles; the dataset of Figure 5.

**Movie M2.** STO N-2_0.mp4: Visualization of the RHEED image stack STO N-2_0.h5 used for the quantitative Auto-RHEED analysis shown in Figure 3. The movie contains 2,923-frame

sequence used to extract time-dependent in-plane lattice spacing, coherence length, streak width, and intensity oscillations.

**Movie M3.** STO N-2_1.mp4: Visualization of the RHEED image stack STO N-2_1.h5 used for the agent-driven Auto-RHEED analysis shown in Figure 4. The movie contains 3,827-frame sequence analyzed through the Auto-RHEED MCP workflow.

**Movie M4.** LSMO5_frames0-1700.mp4: Visualization of the RHEED image stack LSMO5_frames0-1700.h5 used for the AI-inference analysis shown in Figure 6. The movie contains 1,700-frame sequence used for DINOv3 embedding, changepoint analysis, clustering, and YOLO-based diffraction-feature segmentation.

**Data D1.** RHEED_alighnment_STO_110_5 deg_images.npy and RHEED_alighnment_STO_110_5 deg_angles.npy: RHEED frames and corresponding azimuthal angles used for automated SrTiO3 [110] substrate alignment analysis in Figure 2.